\documentclass[aps,twocolumn, amsmath, showpacs, amssymb, prc]{revtex4-1} 
 
\usepackage{graphicx}% Include figure files 
\usepackage{dcolumn}% Align table columns on decimal point 
\usepackage{bm}% bold math 
\begin{document}

\title{Influence of a temperature-dependent shear viscosity
on the azimuthal asymmetries of transverse momentum
spectra in ultrarelativistic heavy-ion collisions}
 
\author{H.\ Niemi${}^{a, b}$, G.S.\ Denicol${}^{c}$, P.\ Huovinen${}^{c}$, E.\ Moln\'ar${}^{b,d}$,  and D.H.\ Rischke${}^{b,c}$}

\affiliation{$^{a}$Department of Physics, P.O. Box 35 (YFL) FI-40014 University of
Jyv\"askyl\"a, Finland}

\affiliation{$^{b}$Frankfurt Institute for Advanced Studies,
Ruth-Moufang-Str.\ 1, D-60438 Frankfurt am Main, Germany}

\affiliation{$^{c}$Institut f\"ur Theoretische Physik, Johann Wolfgang
Goethe-Universit\"at, Max-von-Laue-Str.\ 1, D-60438 Frankfurt am Main, Germany}

\affiliation{$^{d}$MTA Wigner Research Centre for Physics,
H-1525 Budapest, P.O.Box 49, Hungary}

\begin{abstract} 

We study the influence of a temperature-dependent shear viscosity over entropy density 
ratio $\eta/s$, different shear relaxation times $\tau_\pi$, as well as different initial conditions
on the transverse momentum spectra of charged hadrons and identified
particles. We investigate the azimuthal flow asymmetries as a function of 
both collision energy and centrality.
The elliptic flow coefficient turns out to be dominated by
the hadronic viscosity at RHIC energies. Only at higher collision energies the
impact of the viscosity in the QGP phase is visible in the flow asymmetries.
Nevertheless, the shear viscosity near the QCD transition region 
has the largest impact on the collective flow of the system.
We also find that the centrality dependence of the elliptic flow is sensitive
to the temperature dependence of $\eta/s$. 

\end{abstract}

 
\maketitle 

\section{Introduction}

Determining the properties of the quark-gluon plasma (QGP) is nowadays one of the
most important goals in high-energy nuclear physics.
For a system of weakly interacting particles reliable results
can be obtained from first-principle quantum field-theoretical calculations.
Unfortunately, for strongly interacting matter these tools provide only a 
limited amount of information.
It is, however, possible to calculate the thermodynamical properties
of such matter numerically from the theory of strong interactions, quantum
chromodynamics (QCD). These lattice QCD calculations 
show that if the temperature is sufficiently high, the matter undergoes a 
transition from a confined phase where the relevant degrees of freedom 
are hadrons, to a deconfined phase where the degrees of freedom are
quarks and gluons, the so-called QCD transition~\cite{latticereview}.

In recent years, experiments at the Relativistic Heavy-Ion Collider (RHIC) 
at Brookhaven National Laboratory~\cite{experiments} and the Large Hadron 
Collider (LHC) at CERN have provided a wealth of data from which one
could in principle obtain information about the QGP.
However, to compare these data with lattice QCD results is not straightforward. 
So far, lattice calculations have provided reliable results for static thermodynamical 
properties of QCD matter, e.g.\ the equation of state (EoS). The system created
in heavy-ion collisions is, however, not static but dynamical, because
it expands and cools in a very short time span of order $10^{-23}$ seconds.
Obviously, in order to be able to properly interpret the
experimental results and infer the properties of QCD matter, 
we also need a good understanding of the dynamics
of heavy-ion collisions. 

Fluid dynamics is one of the most commonly used frameworks
to describe the space-time evolution of the created fireball,
because the complicated microscopic dynamics of the matter
is encoded in only a few macroscopic parameters like the
EoS and the transport coefficients. 

Currently, fluid-dynamical models give a reasonably good quantitative 
description of transverse momentum spectra of 
hadrons and their centrality dependence~\cite{Huovinen:2006jp, allviscous,
SongBassetal, Shen:2010uy, Song:2010mg}. So far, most calculations assume that 
the shear viscosity to entropy density ratio $\eta/s$ is constant, and
they show that, in order to describe the azimuthal asymmetries of the spectra, e.g.\
the elliptic flow coefficient $v_2$, this constant must be very small,
of order $0.1$. However, for real physical systems,
$\eta/s$ depends (at least) on the temperature \cite{Csernai:2006zz}. 
A constant value of $\eta/s$ can only be justified as an average 
over the space-time evolution of the system. It is not clear 
how this average is related to the temperature dependence of $\eta/s$.

In previous work \cite{Niemi:2011ix, SQM}, we have 
studied the consequences of relaxing the
assumption of a constant $\eta/s$. We 
found that the relevant temperature region where the shear 
viscosity affects the elliptic flow most varies with the collision
energy. At RHIC the most relevant region is around and below the
QCD transition temperature, while for higher collision energies
the temperature region above the transition becomes more and more
important. In this work we shall extend our previous study and provide 
a more detailed picture of the temperature regions that affect 
elliptic flow as well as higher harmonics at a given collision energy.

This paper is organized in the following way. In Sec.\ II, we describe our
fluid-dynamical framework and its numerical implementation.
In Sec.\ III, we specify the EoS, the transport coefficients,
and the initialization. Sections IV and V contain a detailed compilation
of our results, some of which were already shown in Refs.~\cite{Niemi:2011ix, SQM}.
We present the transverse momentum spectra and the elliptic flow of hadrons 
at various centralities with different parameterizations of $\eta/s$ as function
of temperature. We also study the impact of different initial conditions and
of the choice of the relaxation time for the shear-stress tensor.
In Sec.\ VI, we investigate evolution of the elliptic flow in more detail
and, in Sec.\ VII, find the temperature regions where $v_2$ and $v_4$ are most 
sensitive to the value of $\eta/s$. Finally, we summarize our results and give
some conclusions. We use natural units $\hbar = c = k = 1$ throughout the paper.

\section{Fluid dynamics}

\subsection{Formalism}

In order to describe the evolution of a system
on length scales much larger than a typical microscopic 
scale, for instance the mean-free path, it is sufficient to characterize the
state of matter by a few macroscopic fields, namely the energy-momentum tensor $T^{\mu\nu}$
and, possibly, some charge currents $N^{\mu}_a$.
Fluid dynamics is equivalent to the local conservation laws
for these fields,
\begin{equation}
\partial_\mu T^{\mu\nu} = 0\,, \qquad \partial_\mu N^{\mu}_a = 0 \, .
\label{eq:conservation}
\end{equation}

In the absence of conserved charges and bulk viscosity, the 
energy-momentum tensor $T^{\mu\nu}$ can be decomposed as
\begin{equation}
 T^{\mu\nu} = e u^\mu u^\nu - P \Delta^{\mu\nu} + \pi^{\mu\nu},
\end{equation}
where $u^{\mu} = T^{\mu \nu} u_\nu/e$ is the fluid four-velocity, 
$e$ is the energy density in the local rest frame of the fluid, i.e., in the
frame where $u^{\mu} = (1, 0, 0, 0)$, and 
$P$ is the thermodynamic pressure.
The shear-stress tensor is defined as $\pi^{\mu\nu} = T^{\langle \mu \nu \rangle}$,
where the angular brackets $\left<\right>$ denote the symmetric and traceless part of 
the tensor orthogonal to the fluid velocity.
With the $(+,-,-,-)$ convention for the metric tensor $g^{\mu\nu}$,
the projector $\Delta^{\mu\nu} = g^{\mu\nu} - u^\mu u^\nu$. 

If the system is sufficiently close to local thermodynamical
equilibrium, the energy-momentum conservation equations can be closed by providing
the EoS, $P(T)$, the equations determining $\pi^{\mu\nu}$,
and the transport coefficients entering these equations, e.g.\
the shear viscosity $\eta(T)$. The EoS $P(T)$ and the shear 
viscosity $\eta(T)$ can in principle be computed by integrating 
out the dynamics on microscopic length scales.

While the conservation laws are exact for any system, the 
equations determining the shear-stress tensor require certain 
approximations, so that the only variables entering the equations of
motion are those that appear in the energy-momentum tensor, namely
$e$, $u^{\mu}$, and $\pi^{\mu\nu}$. 
In the so-called relativistic Navier-Stokes approximation,
the shear-stress tensor is directly proportional to the gradients
of the four-velocity,
\begin{equation}
 \pi^{\mu\nu} = 2\eta \sigma^{\mu \nu} \equiv 2\eta \partial^{\left<\mu\right.}u^{\left.\nu\right>}.
\label{eq:NS}
\end{equation}
We note that in this approximation the shear-stress tensor 
is not an independent dynamical variable.

Unfortunately, this approximation results in parabolic
equations of motion, and subsequently the signal speed is not 
limited in this theory. In relativistic fluid dynamics 
this violation of causality leads to the existence of linearly 
unstable modes, which make relativistic Navier-Stokes (NS) 
theory useless for practical applications~\cite{his, Shipu}.

A commonly used approach that cures these instability and acausality 
problems is Israel-Stewart (IS) theory \cite{IS}. In this 
approach the shear-stress tensor, the heat flow and bulk viscous pressure 
are introduced as independent dynamical variables and fulfill coupled, 
so-called relaxation-type differential equations of motion.
Assuming vanishing heat-flow and bulk viscosity, the relaxation equation for 
the shear-stress tensor can be written as \cite{resTFD}, 
\begin{eqnarray}
\notag
 \tau_\pi \dot{\pi}^{\left< \mu\nu \right>} &+& \pi^{\mu\nu}   = 2\eta \sigma^{\mu \nu} + \lambda_1 \pi^{\mu\nu} \theta  
 + \lambda_2 \sigma^{\left< \mu\right.}_{\,\,\,\, \alpha} \pi^{\left. \nu\right> \alpha} \\
   &   + &\lambda_3 \pi^{\left< \mu\right.}_{\,\,\,\, \alpha} \pi^{\left. \nu\right> \alpha}
      + \lambda_4 \omega^{\left< \mu\right.}_{\,\,\,\, \alpha} \pi^{\left. \nu\right> \alpha},
\label{eq:IS}
\end{eqnarray}
where $\dot{A} = u^\mu\partial_\mu A$ denotes the 
comoving derivative of $A$ and $\theta = \partial_\mu u^\mu$
is the expansion scalar. The shear-relaxation time $\tau_\pi$ 
is the slowest time scale of the underlying microscopic theory
\cite{paperpoles}. Formally, IS theory can be derived by neglecting 
all faster microscopic time scales~\cite{resTFD}.
Like $\tau_\pi$, the coefficients $\lambda_i$ can in principle be calculated from the 
underlying microscopic theory, i.e., in our case QCD.
Unfortunately, for QCD the transport coefficients appearing in Eq.~\eqref{eq:IS} are still
largely unknown. For the sake of simplicity, in this work we use
$\lambda_1 = -4/3$, obtained from the Boltzmann equation for a massless gas~\cite{IS},
and $\lambda_2 = \lambda_3 = \lambda_4 = 0$.
The shear-relaxation time and the shear viscosity are left as free parameters.

Instead of the full (3+1)--dimensional treatment we
consider a simplified evolution where the expansion in
the $z$-direction is described by boost-invariant scaling flow~\cite{Bjorken:1982qr}, 
i.e., the longitudinal velocity is given by $v_z = z/t$, and the scalar densities 
are independent of the
space-time rapidity $\eta_s = \frac{1}{2} \log\left(\frac{t+z}{t-z}\right)$. Here, $t$ is 
the time measured in laboratory coordinates. In this approximation the full 
evolution depends only on the coordinates $(\tau, x, y)$, where $x$ and $y$ are 
the transverse coordinates and $\tau = \sqrt{t^2-z^2}$ is the longitudinal
proper time.

\subsection{Numerical implementation}

Once the initial values of the components of the energy-momentum
tensor are specified at a given initial time $\tau_0$, the space-time evolution of the system is obtained
by solving the conservation laws \eqref{eq:conservation} together with the
IS equations \eqref{eq:IS}. 

The conservation laws are solved using the algorithm developed in Refs.~\cite{Borisetal}
and generalized to more than one dimension in Ref.~\cite{Zalesak}. This method,
known as SHASTA for ''SHarp and Smooth Transport Algorithm'', solves
equations of the type
\begin{equation}
\partial_t U + \partial_i (v_i U) = S(t,\mathbf{x}) \, ,
\label{eq:generic}
\end{equation}
where $U = U(t,\mathbf{x})$ is for example $T^{00}$, $T^{0i}$, $\ldots$, $v_i$ is
the $i$th component of three-velocity, and $S(t,\mathbf{x})$ is a
source term, for more details see Ref.~\cite{Molnar:2009tx}.

We can further stabilize SHASTA by letting the antidiffusion
coefficient $A_{ad}$ which controls the amount of numerical 
diffusion to be proportional to
\begin{equation}
 \frac{1}{\left(k/e\right)^2 + 1},
\end{equation}
where $e$ is the energy density in the local rest frame, and
$k$ is some constant of order $10^{-5}$ GeV/fm$^3$. In 
this way, $A_{ad}$ goes smoothly to zero near the boundaries 
of the grid, i.e., we increase the amount of numerical diffusion in that 
region. We have checked that this neither affects the 
solution nor produces more entropy inside the decoupling surface.

The relaxation equation \eqref{eq:IS} could also be
solved using SHASTA. However, we noticed that solving
it by replacing the spatial gradient at grid point $i$ 
on the left-hand side of Eq.~\eqref{eq:IS}
by a centered second-order difference, 
\begin{equation}
 \partial_x U_i = \frac{U_{i+1} - U_{i-1}}{2\Delta x},
\label{eq:2ndorder}
\end{equation}
where $U=\pi^{\mu\nu}$, yields a more stable algorithm. Time derivatives in the source
terms are simply taken as first-order backward differences.
Like in SHASTA, all spatial gradients in the source terms
are discretizised according to Eq.~\eqref{eq:2ndorder}.

\subsection{Freeze-out}

We assume that freeze-out, i.e., the transition from the
fluid-dynamical system to free-streaming particles happens
on a hypersurface of constant temperature. Unless otherwise stated,
we assume that the freeze-out temperature is $T_{\rm dec} = 100$ MeV.
We include all 2- and 3-particle decays of 
hadronic resonances according to Ref.~\cite{Sollfrank:1991xm}.

The transverse momentum distribution of hadrons is calculated 
using the Cooper-Frye description~\cite{Cooper:1974mv}. For the 
final spectra we need to know the local single-particle momentum distribution 
functions of hadrons on the freeze-out surface. Here, we employ
the widely used 14-moment ansatz where the correction to
the local-equilibrium distribution 
$f_{0i} = \left\{\exp\left[\left(u_\mu p_i^\mu-\mu_i\right)/T\right] \pm 1 \right\}^{-1}$
of a hadron of species $i$ with four-momentum $p_i^\mu$ 
is given by \cite{Teaney:2003kp}
\begin{equation}
 \delta f_i = f_{0i} \frac{p_i^\mu p_i^\nu \pi_{\mu\nu}}{T^2\left(e+P\right)}.
\label{eq:deltaf}
\end{equation}
We note that this functional form for $\delta f$ is merely an ansatz.
If dissipative fluid dynamics is derived from the Boltzmann equation
without assuming the 14-moment approximation, 
the full expansion of $\delta f$ contains an infinite number 
of terms, for details see Ref.~\cite{resTFD}. The effect of 
this  will be studied in a future work.

\section{Parameters}

\subsection{Equation of State}

As EoS we use the recent $s95p$-PCE-v1 parameterization of lattice QCD 
results~\cite{Huovinen:2009yb}. In this
parameterization, the high-temperature part is matched to 
recent results of the hotQCD collaboration~\cite{Cheng:2007jq,Bazavov:2009zn} 
and smoothly connected to the low-temperature part described as a hadron resonance gas.  
All hadrons listed in Ref.~\cite{pdg} up to a mass of 2 GeV are included in the
hadronic part of the EoS. The system is assumed to chemically
freeze-out at $T_\mathrm{chem}=150$ MeV. Below this temperature the
EoS is constructed according to Refs.~\cite{Bebie:1991ij, Hirano:2002ds, Huovinen:2007xh}.
This construction assumes that the evolution below  $T_{\rm chem}$ is isentropic. 
Strictly speaking this is not the case in viscous
hydrodynamics since dissipation causes an increase in
entropy. However, we have checked that in our calculations the viscous
entropy production from all fluid cells with temperatures below
$T_\mathrm{chem} = 150$\,MeV is less than 1\% of the initial entropy, whereas the entropy
production during the entire evolution ranges from $3-14$ \%, depending
on the collision energy and the $\eta/s$ parameterization.

\subsection{Transport coefficients}

The temperature-dependent shear viscosity is parametrized as follows.
In all cases, we take the minimum of $\eta/s$ to be at 
$T_{\rm tr} = 180$ MeV. Unless otherwise stated, the value of $\eta/s$ 
at the minimum is assumed to equal the lower bound $\eta/s = 0.08$ conjectured
in the framework of the AdS/CFT correspondence~\cite{AdSCFT}. 

%%%%%%%%%%%%%%%%%%%%% FIGURE %%%%%%%%%%%%%%%%%%%%%%%%%%%%%%%%
\begin{figure}[bht] 
% \vspace{-0.5cm} 
\includegraphics[width=8cm]{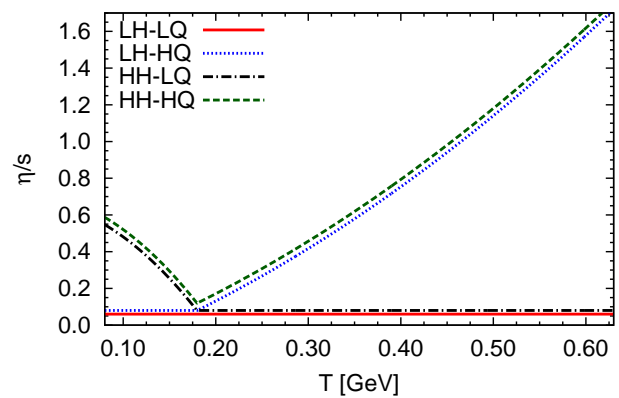} 
% \vspace{-0.3cm} 
\caption{\protect\small (Color online) Different parameterizations
of $\eta/s$ as a function of temperature. The \emph{(LH-LQ)} line
is shifted downwards and the \emph{(HH-HQ)} line 
upwards for better visibility.}
% \vspace{-0.3cm} 
\label{fig:eta}
\end{figure} 
%%%%%%%%%%%%%%%%%%%%% FIGURE %%%%%%%%%%%%%%%%%%%%%%%%%%%%%%%% 
The parameterization of the hadronic viscosity is based on Ref.~\cite{NoronhaHostler:2008ju}
where the authors consider a hadron resonance gas with additional Hagedorn states.
In practice, we use a temperature dependence of $\eta/s$ of
the following functional form~\cite{Niemi:2011ix, Denicol:2010tr},
\begin{equation}
\label{lowT}
 \frac{\eta}{s}\Big|_{\rm HRG} = 0.681 - 0.0594\, \frac{T}{T_{\rm tr}} - 0.544 \left(\frac{T}{T_{\rm tr}}\right)^2.
\end{equation}
At $T = 100$ MeV this coincides with the $\eta/s$ value given in Ref.~\cite{NoronhaHostler:2008ju}, and
decreases smoothly to the minimum value $\eta/s = 0.08$ at $T_{tr}$. We note
that many authors obtain considerably larger values for the shear viscosity
of hadronic matter, see e.g.\ Refs.~\cite{hadronicviscosity}. Our motivation here is 
to illustrate the effects of hadronic viscosity rather than to use a parameterization
that is as realistic as possible. We shall see that even this low $\eta/s$ leads
to considerable effects for hadronic observables in Au + Au collisions
at RHIC. We further note that, since we are considering a chemically frozen hadron resonance gas below $T_{\rm chem}$, 
while in Ref.~\cite{NoronhaHostler:2008ju} chemical equilibrium is assumed at all temperatures,
the entropy densities, and therefore the values of $\eta$, differ between the two calculations at a
given value of $T < T_{\rm chem}$. 

The high-temperature QGP viscosity is parametrized according to lattice QCD 
results \cite{Nakamura:2004sy} in such a way that it connects to the
minimum of $\eta/s$ at $T_{\rm tr}$. The functional form used is
\begin{equation}
\label{highT}
  \frac{\eta}{s}\Big\rvert_{\rm QGP} = -0.289 + 0.288\, \frac{T}{T_{\rm tr}} + 0.0818 \left(\frac{T}{T_{\rm tr}}\right)^2.
\end{equation}

We take the following four parameterizations of the shear viscosity:
\begin{itemize}
 \item \emph{(LH-LQ)} $\eta/s = 0.08$ for all temperatures,
 \item \emph{(LH-HQ)} $\eta/s = 0.08$ in the hadron gas, and above
	$T = 180$ MeV $\eta/s$ increases according to Eq.\ (\ref{highT}),
 \item  \emph{(HH-LQ)} below $T=180$ MeV, $\eta/s$ is given by
Eq.\ (\ref{lowT}),
and above we set $\eta/s = 0.08$,
 \item \emph{(HH-HQ)} we use Eqs.\ (\ref{lowT}) and (\ref{highT}) for 
the hadron gas and the QGP, respectively.
\end{itemize}
These parameterizations are shown in Fig.~\ref{fig:eta}.
Besides these four cases we also study the effect of varying the value of the minimum 
of $\eta/s$, see Secs.\ \ref{effects_init} and \ref{probing}. 
%%%%%%%%%%%%%%%%%%%%% FIGURE %%%%%%%%%%%%%%%%%%%%%%%%%%%%%%%%
\begin{figure}[bht] 
% \vspace{-0.5cm} 
\includegraphics[width=8cm]{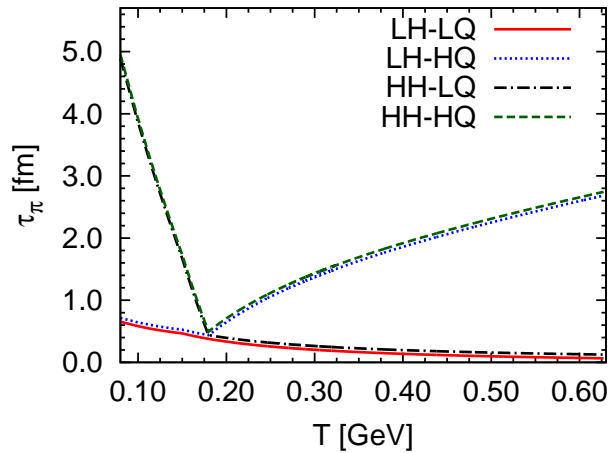} 
\vspace{-0.3cm} 
\caption{\protect\small (Color online) 
Relaxation times corresponding to the different parameterizations of $\eta/s$, for $c_\tau=5$.
The \emph{(LH-LQ)} line
is shifted downwards and the \emph{(HH-HQ)} line 
upwards for better visibility.}
\vspace{-0.3cm} 
\label{fig:tau_relax}
\end{figure} 
%%%%%%%%%%%%%%%%%%%%% FIGURE %%%%%%%%%%%%%%%%%%%%%%%%%%%%%%%% 

In order to complete the description, we also need to specify the relaxation time.
In this work we use a functional form suggested by kinetic theory,
\begin{equation}
 \tau_\pi = c_\tau \frac{\eta}{e+p},
\label{eq:relax}
\end{equation}
where $c_\tau$ is a constant. Causality requires that $c_\tau \geq 2$~\cite{Shipu}. Unless otherwise
stated, we shall use the value $c_\tau = 5$ which coincides with the value obtained from the 
Boltzmann equation in the 14-moment approximation for a massless gas of classical
particles \cite{Denicol:2010xn}. The relaxation times 
corresponding to the parameterizations above are shown in Fig.~\ref{fig:tau_relax}.
The effect of varying the relaxation time separately 
from $\eta$ is also studied in Sec.~\ref{effects_init}.

\subsection{Initial state}

We still need to specify the initial state at some proper time $\tau_0$. For a boost-invariant
system it is sufficient to provide the components of the energy-momentum
tensor in the transverse plane at $z=0$, i.e., $\eta_s = 0$.
Within our approximations these are the local energy density, the
initial transverse velocity, and the three independent
components of the shear-stress tensor. Here, we will assume that
the initial transverse velocity is zero and, unless otherwise
stated, the initial shear-stress tensor is also assumed to be zero.

For the initial time we choose $\tau_0 = 1$ fm.
The energy density $e(\tau_0, x, y)$ is based on the optical 
Glauber model by assuming that the energy density is a function of 
the density of binary nucleon-nucleon collisions $n_{\rm BC}$, 
or the density of wounded nucleons $n_{\rm WN}$, or both,
\begin{equation}
 e(\tau_0, x, y) = C_e f(n_{\rm BC}, n_{\rm WN}).
\end{equation}
The overall normalization, $C_e$, is fixed in order to reproduce the observed
multiplicities in the most central $\sqrt{s_{NN}} = 200$ GeV Au+Au collisions 
at RHIC, and in $\sqrt{s_{NN}} = 2.76$ GeV Pb+Pb collisions 
at LHC.

The centrality dependence of the multiplicity is reproduced in this work in two
different ways:
\begin{itemize}
 \item \emph{BCfit}: choosing $f$ to be a polynomial in $n_{\rm BC}$,
\begin{equation}
f(n_{\rm BC}) = n_{\rm BC} + c_1 n_{\rm BC}^2 + c_2 n_{\rm BC}^3.
\label{eq:BCfit}
\end{equation}
\item \emph{GLmix}: using a superposition of $n_{\rm BC}$ and $n_{\rm WN}$,
\begin{equation}
f(n_{\rm BC}, n_{\rm WN}) = d_1 n_{\rm BC} + (1 - d_1) n_{\rm WN}.
\label{eq:GLmix}
 \end{equation}
\end{itemize}
Here, the coefficient $c_2$ is introduced in order to guarantee that the 
parameterizations are monotonically increasing
with increasing binary-collision or wounded-nucleon density. This ensures
that the highest energy density is in the center of the system, i.e.,
at $x=y=0$. 

For a given impact parameter, the optical Glauber model yields a different 
number of participants and different centrality classes than the Monte Carlo 
Glauber models commonly used by the experimental collaborations. 
Using the optical Glauber model, we can either choose to
reproduce the multiplicity as a function of the number 
of participants or as a function of centrality classes. In general, this leads to different
coefficients $c_i$ and $d_1$. 
Here, we choose to determine the initial conditions by requiring that
the centrality dependence of the charged particle multiplicity as a function
of the number of participants~\cite{Adler:2003cb, Aamodt:2010cz} is reproduced. 
We have checked that, if we determine the centrality dependence by matching to the centrality
classes given by the optical Glauber model, the elliptic flow is more suppressed
in central and enhanced in peripheral collisions at RHIC energies, while at LHC
energies it remains practically unchanged.
In order to be fully consistent
with the experimental determination of the centrality classes,
one would need to generate fluctuating initial conditions via
a Monte Carlo Glauber model, see e.g. Refs.~\cite{Holopainen:2010gz, Schenke:2011bn}.

For $\sqrt{s_{NN}} = 5.5$ TeV Pb+Pb collisions we use the multiplicity
in the most central collisions as predicted by the EKRT model~\cite{Eskola:2005ue}.
In this case the centrality dependence is assumed to follow binary scaling,
i.e., $c_1 = c_2 = 0$ in Eq.~\eqref{eq:BCfit}. All initialization parameters
are shown in Table~\ref{tab:ini}.
\begin{table}[t]
\begin{center}
\begin{tabular}{|c|c|c||c||c|}
\hline
$\sqrt{s_{NN}}$ [GeV] & $c_1$ [fm$^{-2}$] & $c_2$ [fm$^{-4}$] & $d_1$ & $T_{\rm max}$ [MeV]  \\ \hline
200	& $-0.032$	& $0.00035$	& $0.1$  & $313$	\\ \hline
2760	& $-0.01$	& $0.0001$	& $0.7$  & $430$	\\ \hline
5500	& $0$		& $0$		& $1.0$  & $504$	\\ \hline
\end{tabular}%
\end{center}
\vspace*{-0.1cm}
\caption{\protect\small Initialization parameters for different collision 
energies. The maximum temperature $T_{\rm max}$ is given for the
\emph{BCfit} initialization with the \emph{(LH-LQ)} 
parameterization of $\eta/s$.
For the other initializations $T_{max}$ differs less than 5\%.}
\vspace*{-0.3cm}
\label{tab:ini}
\end{table}

Different parameterizations of $\eta/s$ lead to different
entropy production and therefore different final multiplicity, even 
if the initial state is kept the same. This is especially true for
different parameterizations of the high-temperature shear viscosity, since most
of the entropy is produced during the early stages of the collision \cite{Dumitru:2007qr}. We compensate
this using different overall normalizations e.g.\ between the \emph{(HH-LQ)} and 
\emph{(HH-HQ)} parameterizations. Entropy production during the hadronic evolution 
is small and not compensated. The centrality dependence of the entropy production is also 
different for different $\eta/s$ parameterizations. Since it leads to at most a 5\%
difference in the final multiplicities and is hardly visible in the results,
it is not corrected here.

\section{Results and comparison with experimental data}

In this section we use the initializations and parameterizations of
$\eta/s$ given above, and compare the results with experimental
data from RHIC and LHC.

\subsection{Transverse momentum spectra and elliptic flow at RHIC}

\begin{figure}[bht] 
% \vspace{-0.5cm} 
\includegraphics[width=8.0cm]{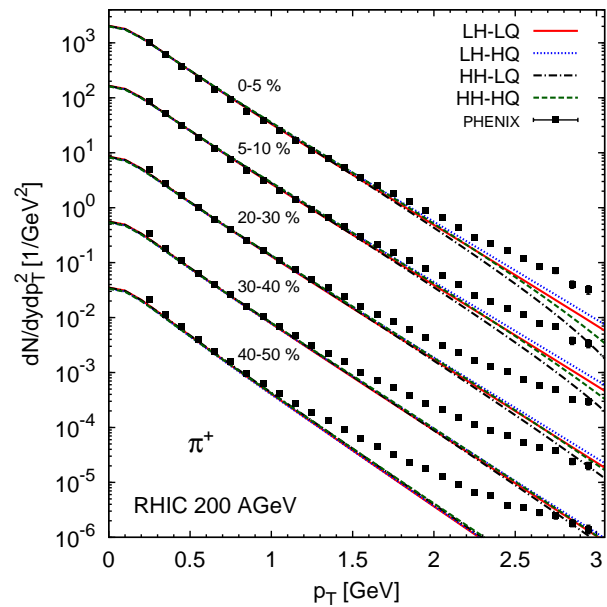} 
% \vspace{-0.3cm} 
\caption{\protect\small (Color online) Pion spectra at RHIC, with \emph{BCfit} initialization.}
% \vspace{-0.3cm} 
\label{fig:pionspectrum_BCfit}
\end{figure}

In Fig.~\ref{fig:pionspectrum_BCfit} we show the $p_T$-spectra of
pions for different centrality classes for RHIC $\sqrt{s_{NN}} = 200$ GeV
Au+Au collisions and compare them with PHENIX data~\cite{Adler:2003cb}.
We only show results using the \emph{BCfit} initialization; those for
the \emph{GLmix} initialization are very similar. The freeze-out temperature is
chosen as $T_{\rm dec} = 100$ MeV. This choice reproduces the slopes of the 
$p_T$-spectra quite well. 

Once we correct the normalization of the initial energy density 
profile for different entropy production,
the slopes of the $p_T$-spectra are practically unaffected by the
$\eta/s$ parameterizations. We note that in our earlier work~\cite{Niemi:2011ix}
this correction was not made, and the different $\eta/s$ parameterizations
lead not only to different multiplicities but also to different slopes 
for the $p_T$-spectra. 
This effect was even more pronounced at LHC than at RHIC, due to an 
increase in entropy production
caused by larger gradients appearing with an earlier initialization 
time $\tau_0 = 0.6$ fm. 

%%%%%%%%%%%%%%%%%%%%% FIGURE %%%%%%%%%%%%%%%%%%%%%%%%%%%%%%%%
%%%%%%%%%%%%%%%%%%%%% FIGURE %%%%%%%%%%%%%%%%%%%%%%%%%%%%%%%% 

\begin{figure}[bht] 
% \vspace{-0.5cm} 
\includegraphics[width=8.0cm]{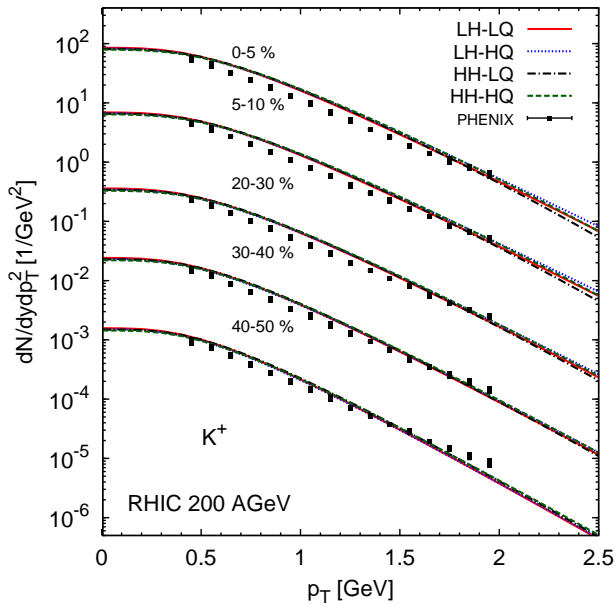} 
% \vspace{-0.3cm} 
\caption{\protect\small (Color online) Kaon spectra at RHIC, with \emph{BCfit} initialization.}
\vspace{-0.3cm} 
\label{fig:kaonspectrum_BCfit}
\end{figure}

\begin{figure}[bht] 
% \vspace{-0.5cm} 
\includegraphics[width=8.0cm]{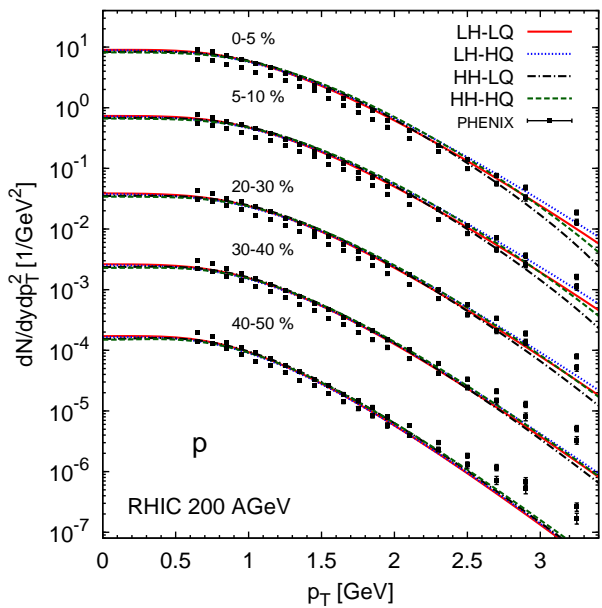} 
% \vspace{-0.3cm} 
\caption{\protect\small (Color online) Proton spectra at RHIC, with \emph{BCfit} initialization.}
\vspace{-0.3cm} 
\label{fig:protonspectrum_BCfit}
\end{figure}

The kaon spectra are shown in Fig.~\ref{fig:kaonspectrum_BCfit} and 
the proton spectra in Fig.~\ref{fig:protonspectrum_BCfit} with
the \emph{BCfit} initialization. Both are compared
with PHENIX data~\cite{Adler:2003cb}. Because we do not consider
net-baryon number in our calculations, the proton and anti-proton
spectra are identical. For this reason we show both the proton and 
the anti-proton data in Fig.~\ref{fig:protonspectrum_BCfit}.

For both kaons and protons the calculated spectra are slightly
more curved than the data and they also lie above 
the data. As for the pions, the slopes of the spectra are
practically independent of the $\eta/s$ parameterization.
%%%%%%%%%%%%%%%%%%%%% FIGURE %%%%%%%%%%%%%%%%%%%%%%%%%%%%%%%%
%%%%%%%%%%%%%%%%%%%%% FIGURE %%%%%%%%%%%%%%%%%%%%%%%%%%%%%%%% 
%%%%%%%%%%%%%%%%%%%%% FIGURE %%%%%%%%%%%%%%%%%%%%%%%%%%%%%%%%
%%%%%%%%%%%%%%%%%%%%% FIGURE %%%%%%%%%%%%%%%%%%%%%%%%%%%%%%%% 

\begin{figure}[bht] 
% \vspace{-0.5cm} 
\includegraphics[width=8.5cm]{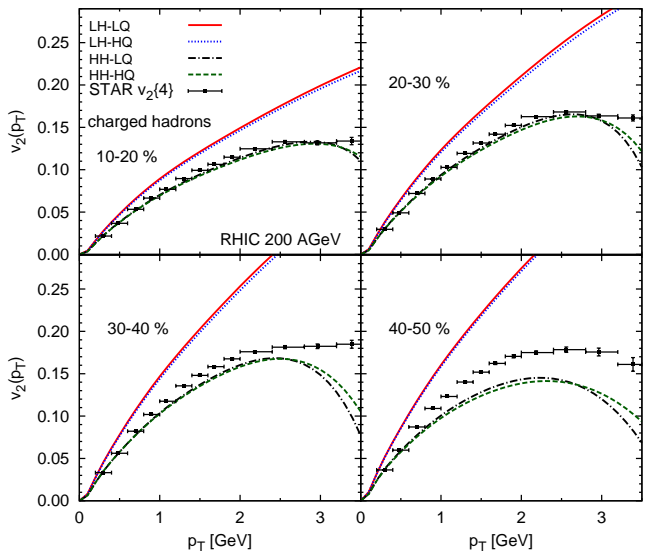} 
% \vspace{-0.3cm} 
\caption{\protect\small (Color online) Charged hadron $v_2(p_T)$ at RHIC, with \emph{BCfit} initialization.}
% \vspace{-0.3cm} 
\label{fig:v2charged_BCfit}
\end{figure}

\begin{figure}[bht] 
% \vspace{-0.5cm} 
\includegraphics[width=8.5cm]{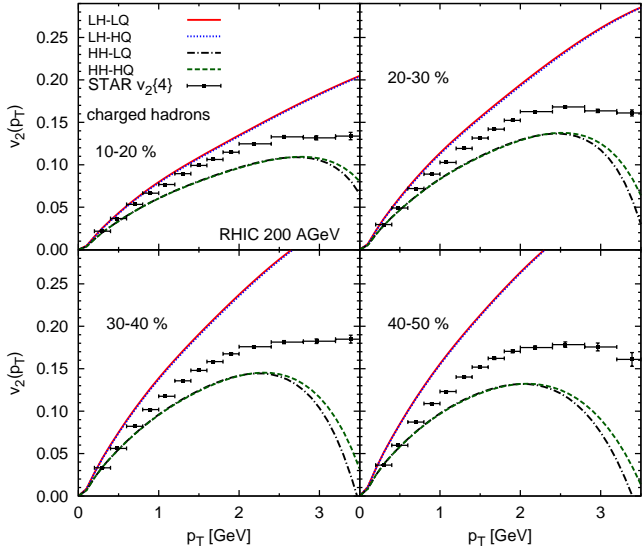} 
% \vspace{-0.3cm} 
\caption{\protect\small (Color online) Charged hadron $v_2(p_T)$ at RHIC, with \emph{GLmix} initialization.}
% \vspace{-0.3cm} 
\label{fig:v2charged_BCmix}
\end{figure}

Figure~\ref{fig:v2charged_BCfit} shows the $p_T$-differential
elliptic flow $v_2(p_T)$ of charged hadrons for different 
centrality classes using the \emph{BCfit} initialization.
Similarly, Fig.~\ref{fig:v2charged_BCmix}
shows the elliptic flow for the \emph{GLmix} initialization.
The calculations are compared with the four-particle cumulant
data from the STAR collaboration~\cite{Bai}. 

As was observed in Ref.~\cite{Niemi:2011ix}, the differential elliptic
flow is largely independent of the high-temperature $\eta/s$
parameterization, but highly sensitive on the hadronic $\eta/s$
at RHIC. This holds for all centrality classes. The suppression
of the elliptic flow due to the hadronic viscosity is even more
enhanced in more peripheral collisions. Note that with the
\emph{BCfit} initialization, the elliptic flow in the most central collision
class is reproduced by the parameterizations with a large hadronic viscosity,
while with the \emph{GLmix} initialization the elliptic flow in the
same centrality class is better described by taking a constant $\eta/s=0.08$.
However, with the latter choice the elliptic flow tends to be
overestimated in more peripheral collisions. On the other
hand, the temperature-dependent hadronic $\eta/s$ gives the centrality
dependence correctly down to the $30-40$ \% centrality class. In 
even more peripheral collisions a large hadronic viscosity tends
to suppress the elliptic flow too much.

\begin{figure}[bht] 
% \vspace{-0.5cm} 
\includegraphics[width=8.5cm]{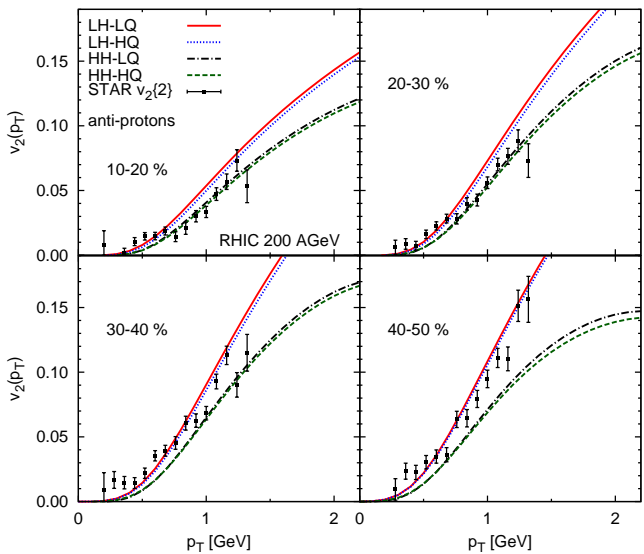} 
% \vspace{-0.3cm} 
\caption{\protect\small (Color online) Proton $v_2(p_T)$ at RHIC, with \emph{BCfit} initialization.}
% \vspace{-0.3cm} 
\label{fig:v2protons_BCfit}
\end{figure}

Figure~\ref{fig:v2protons_BCfit} shows $v_2(p_T)$ for protons
with the \emph{BCfit} initialization compared to the two-particle 
cumulant data from the STAR collaboration~\cite{Adams:2004bi}.
The protons show qualitatively the same response to the different
$\eta/s$ parameterizations as all charged hadrons, i.e., 
$v_2(p_T)$ depends strongly on the hadronic viscosity, but is
almost independent of the high-temperature $\eta/s$. 
Since we use a smooth initialization, with no initial-state 
fluctuations included, quantitative comparisons with two- or
four-particle cumulant data are not straightforward.

%%%%%%%%%%%%%%%%%%%%% FIGURE %%%%%%%%%%%%%%%%%%%%%%%%%%%%%%%%
%%%%%%%%%%%%%%%%%%%%% FIGURE %%%%%%%%%%%%%%%%%%%%%%%%%%%%%%%% 
%%%%%%%%%%%%%%%%%%%%% FIGURE %%%%%%%%%%%%%%%%%%%%%%%%%%%%%%%%
%%%%%%%%%%%%%%%%%%%%% FIGURE %%%%%%%%%%%%%%%%%%%%%%%%%%%%%%%% 
%%%%%%%%%%%%%%%%%%%%% FIGURE %%%%%%%%%%%%%%%%%%%%%%%%%%%%%%%%
%%%%%%%%%%%%%%%%%%%%% FIGURE %%%%%%%%%%%%%%%%%%%%%%%%%%%%%%%% 

\subsection{Transverse momentum spectra and elliptic flow at LHC}

\begin{figure}[bht] 
% \vspace{-0.5cm} 
\includegraphics[width=8.0cm]{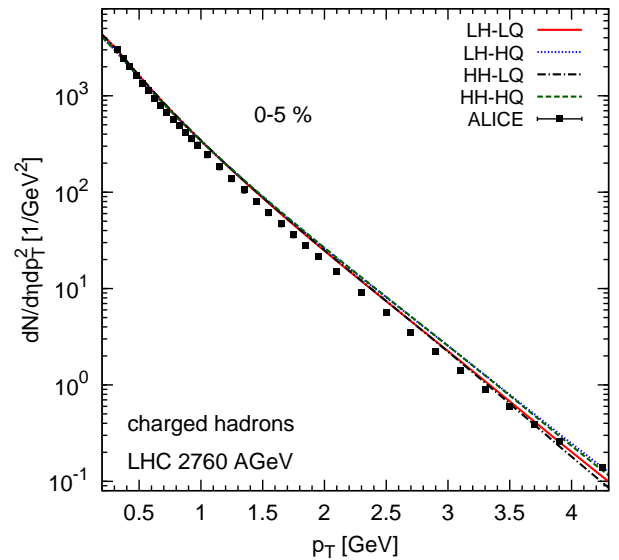} 
% \vspace{-0.3cm} 
\caption{\protect\small (Color online) Charged hadron spectra at LHC, with \emph{BCfit} initialization.}
% \vspace{-0.3cm} 
\label{fig:chargedspectrum_BCfit}
\end{figure}

Transverse momentum spectra of charged hadrons in most central
Pb+Pb collisions with $\sqrt{s_{NN}} = 2.76$ TeV at LHC are
shown in Fig.~\ref{fig:chargedspectrum_BCfit}. At LHC, both
initializations \emph{BCfit} and \emph{GLmix} give very similar results
for both elliptic flow and the spectra, because the contribution
from binary collisions is large, of order $\sim 70$ \%, see Table~\ref{tab:ini}.
Therefore, we show only results with the \emph{BCfit} initialization;
these are compared to data from the ALICE collaboration~\cite{Aamodt:2010jd}. The 
calculated spectra are somewhat flatter than the data. Here, we
have used the same decoupling temperature as at RHIC, 
i.e., $T_{\rm dec} = 100$ MeV. We could improve the agreement
with the data by decoupling at even lower temperature than at RHIC. 
Another way to improve the agreement is choosing a larger chemical 
freeze-out temperature. This would give steeper spectra,
but the proton multiplicity at RHIC would then be overestimated.
However, we have tested that the dependence of the spectra and
the elliptic flow on $\eta/s$ is unchanged by these details.

As was the case at RHIC, at LHC the slopes of the 
spectra are practically independent of the $\eta/s$ parameterization.
We note that here we have used the initialization time $\tau_0 = 1.0$ fm,
i.e., the same as at RHIC. In Ref.~\cite{Niemi:2011ix} we observed a 
quite visible correlation between the shear viscosity and the spectral slopes.
Here, the later initialization time and the fact that we now compensate 
for the entropy production between different $\eta/s$ parameterizations 
almost completely removes this correlation. However, the earlier the
evolution starts, the more the viscosity will affect the slopes.
%%%%%%%%%%%%%%%%%%%%% FIGURE %%%%%%%%%%%%%%%%%%%%%%%%%%%%%%%%
%%%%%%%%%%%%%%%%%%%%% FIGURE %%%%%%%%%%%%%%%%%%%%%%%%%%%%%%%% 

\begin{figure}[bht] 
% \vspace{-0.5cm} 
\includegraphics[width=8.5cm]{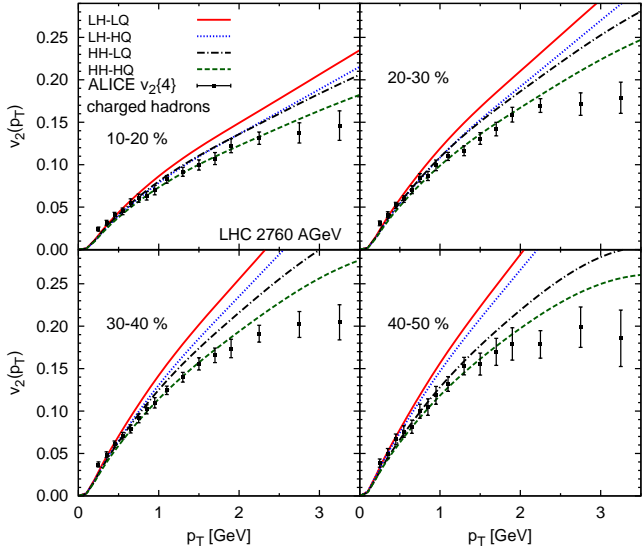} 
% \vspace{-0.3cm} 
\caption{\protect\small (Color online) Charged hadron $v_2(p_T)$ at LHC, with \emph{BCfit} initialization.}
\vspace{-0.3cm} 
\label{fig:v2charged_LHC_BCfit}
\end{figure}

\begin{figure}[bht] 
% \vspace{-0.5cm} 
\includegraphics[width=8.5cm]{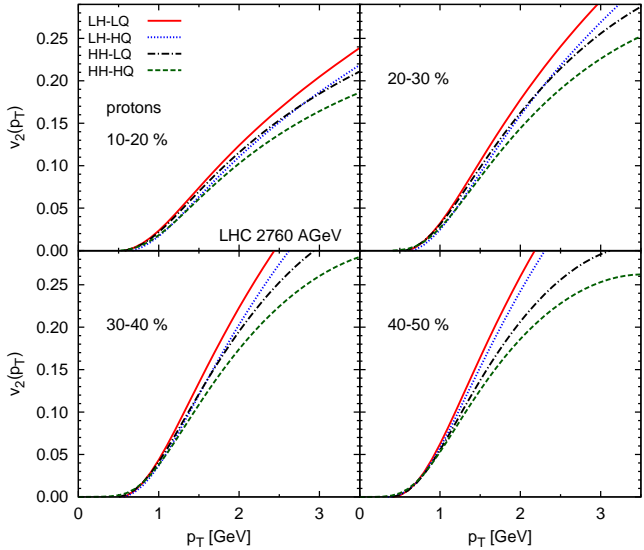} 
% \vspace{-0.3cm} 
\caption{\protect\small (Color online) Proton $v_2(p_T)$ at LHC, with \emph{BCfit} initialization.}
\vspace{-0.3cm} 
\label{fig:v2protons_LHC_BCfit}
\end{figure}

The $p_T$-differential elliptic flow for all charged hadrons
is shown in Fig.~\ref{fig:v2charged_LHC_BCfit} and for protons
in Fig.~\ref{fig:v2protons_LHC_BCfit}. The charged hadron elliptic
flow is compared with ALICE four-particle cumulant data~\cite{Aamodt:2010pa}.
We can see that in the $10-20$ \% centrality class, changing the hadronic $\eta/s$ 
or changing the high-temperature $\eta/s$ has quite a similar impact on the
elliptic flow, e.g.\ the difference between the \emph{LH-LQ} and \emph{LH-HQ} 
and between the \emph{LH-LQ} and \emph{HH-LQ} curves is nearly the same.
However, the more peripheral the collision is, the 
more the viscous suppression is dominated by the hadronic $\eta/s$.
This is confirmed by comparing the grouping of the flow curves in the $40-50$
\% centrality class 
at LHC with that at RHIC, cf.\ Figs.~\ref{fig:v2charged_BCfit} and \ref{fig:v2charged_LHC_BCfit}.
As was the case in Au+Au collisions at RHIC, also here the 
grouping of the curves for the protons is similar to that of all 
charged hadrons, cf.\ Fig.~\ref{fig:v2protons_LHC_BCfit}.

Note that, within our set-up, the best agreement with the ALICE data 
is obtained with the \emph{HH-HQ} parameterization, i.e., with a
temperature-dependent $\eta/s$ in both hadronic and high-temperature
phases. However, in the low-$p_T$ region our
calculations systematically underestimate the elliptic flow in all
centrality classes. As was the case with the $p_T$-spectrum, 
decoupling at a lower temperature and choosing a higher chemical 
freeze-out temperature would improve the agreement, without changing the 
grouping of the elliptic flow curves with the $\eta/s$ parameterizations.
%%%%%%%%%%%%%%%%%%%%% FIGURE %%%%%%%%%%%%%%%%%%%%%%%%%%%%%%%%
%%%%%%%%%%%%%%%%%%%%% FIGURE %%%%%%%%%%%%%%%%%%%%%%%%%%%%%%%% 
%%%%%%%%%%%%%%%%%%%%% FIGURE %%%%%%%%%%%%%%%%%%%%%%%%%%%%%%%%
%%%%%%%%%%%%%%%%%%%%% FIGURE %%%%%%%%%%%%%%%%%%%%%%%%%%%%%%%% 

\begin{figure}[bht] 
% \vspace{-0.5cm} 
\includegraphics[width=8.5cm]{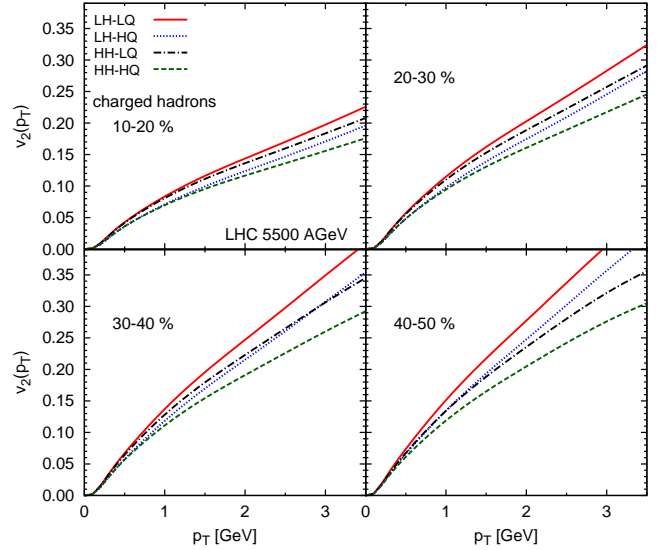} 
% \vspace{-0.3cm} 
\caption{\protect\small (Color online) Charged hadron $v_2(p_T)$ at LHC $5.5$ A TeV, with BC initialization.}
\vspace{-0.3cm} 
\label{fig:v2charged_LHC5500}
\end{figure}

In Fig.~\ref{fig:v2charged_LHC5500} we show the $p_T$-differential
elliptic flow for $\sqrt{s_{NN}} = 5.5$ TeV Pb+Pb collisions. In this
case the viscous suppression of $v_2(p_T)$ is dominated by the high-temperature
$\eta/s$ in central collisions, while peripheral collisions 
resemble more the lower-energy central collisions at LHC, i.e., both
hadronic and high-temperature viscosity contribute similarly to the
suppression. Furthermore, the higher the $p_T$, the more the hadronic
viscosity contributes to the suppression. This happens mainly because 
$\delta f$ increases with both viscosity and $p_T$. 
%%%%%%%%%%%%%%%%%%%%% FIGURE %%%%%%%%%%%%%%%%%%%%%%%%%%%%%%%%
%%%%%%%%%%%%%%%%%%%%% FIGURE %%%%%%%%%%%%%%%%%%%%%%%%%%%%%%%% 

\section{Effects of shear initialization, minimum of \boldmath $\eta/s$ and relaxation time}
\label{effects_init}

One of the main results of Ref.~\cite{Niemi:2011ix} is that, at RHIC, the high-temperature
shear viscosity has very little effect on the elliptic flow. In this section
we elaborate more on this analysis, and explicitly show that this statement
holds for an out-of-equilibrium initialization of the shear-stress tensor as well. 
We also study the effect of varying the relaxation time.

%%%%%%%%%%%%%%%%%%%%% FIGURE %%%%%%%%%%%%%%%%%%%%%%%%%%%%%%%%
\begin{figure}[bht] 
% \vspace{-0.5cm} 
\includegraphics[width=8.0cm]{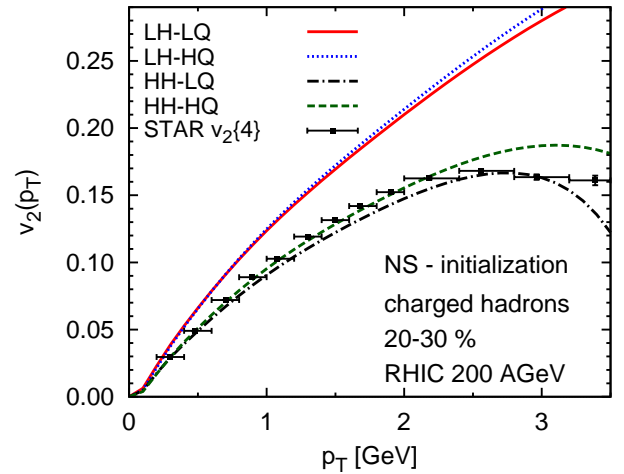} 
% \vspace{-0.3cm} 
\caption{\protect\small (Color online) Charged hadron $v_2(p_T)$ at RHIC, with \emph{BCfit} and NS initialization.}
\vspace{-0.3cm} 
\label{fig:v2charged_RHIC_NS}
\end{figure}
%%%%%%%%%%%%%%%%%%%%% FIGURE %%%%%%%%%%%%%%%%%%%%%%%%%%%%%%%% 
Figure \ref{fig:v2charged_RHIC_NS} shows the elliptic flow of charged 
hadrons in the $20-30$ \% centrality class at RHIC. Instead of 
setting $\pi^{\mu\nu}$ to zero initially, here the so-called Navier-Stokes 
(NS) initialization where the initial
values of the shear-stress tensor are given by the first-order, asymptotic solution
of IS theory, Eq.~\eqref{eq:NS}.
For all $\eta/s$ parameterizations, the NS initialization increases
the entropy production (up to $30$ \%), especially for the parameterizations with
a large high-temperature viscosity. This is corrected by adjusting 
the initial energy density to produce approximately the same final multiplicity.
Although for the parameterizations with a large hadronic $\eta/s$ the 
different shear initializations give slightly different $v_2(p_T)$ curves, 
the grouping of these curves remains intact. 
We emphasize that the NS initialization gives very different 
initial conditions for each viscosity parameterization. If we  
use the same non-zero initial shear stress, e.g.\ $\pi^{\mu\nu} = const. \times \sigma^{\mu\nu}$, 
for each parameterization, the resulting $v_2(p_T)$ curves in each group in 
Fig.~\ref{fig:v2charged_RHIC_NS} would be even closer to each other. 

The NS initialization with a constant $\eta/s=0.08$ has a relatively short 
relaxation time, see Fig.~\ref{fig:tau_relax}. Hence for $\tau_\pi \ll \tau_0$ 
the NS initialization is not a completely unrealistic assumption for the initial values of $\pi^{\mu\nu}$.
However, for larger values of $\eta/s$ the relaxation times are considerably 
larger, $\tau_\pi \gtrsim \tau_0$,
and there is no reason to assume that the asymptotic solution 
could have been reached already at very early times. 

%%%%%%%%%%%%%%%%%%%%% FIGURE %%%%%%%%%%%%%%%%%%%%%%%%%%%%%%%%
\begin{figure}[bht] 
% \vspace{-0.5cm} 
\includegraphics[width=8cm]{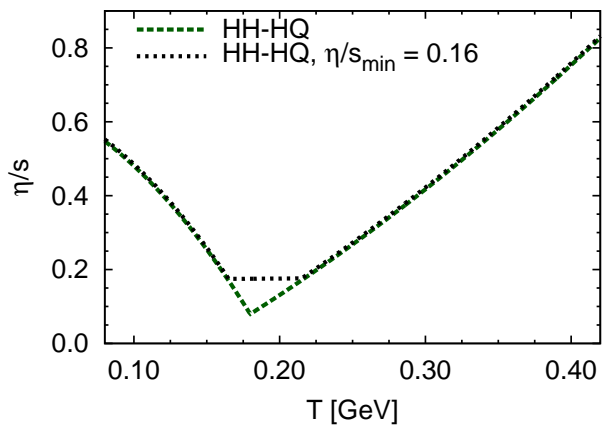} 
% \vspace{-0.3cm} 
\caption{\protect\small (Color online) Parameterizations
of $\eta/s$ as a function of temperature. The \emph{(HH-HQ)} line
the same as in Fig.~\ref{fig:eta}. }
\vspace{-0.3cm} 
\label{fig:etahighmin}
\end{figure} 
%%%%%%%%%%%%%%%%%%%%% FIGURE %%%%%%%%%%%%%%%%%%%%%%%%%%%%%%%% 

\begin{figure}[bht] 
% \vspace{-0.5cm} 
\includegraphics[width=8.0cm]{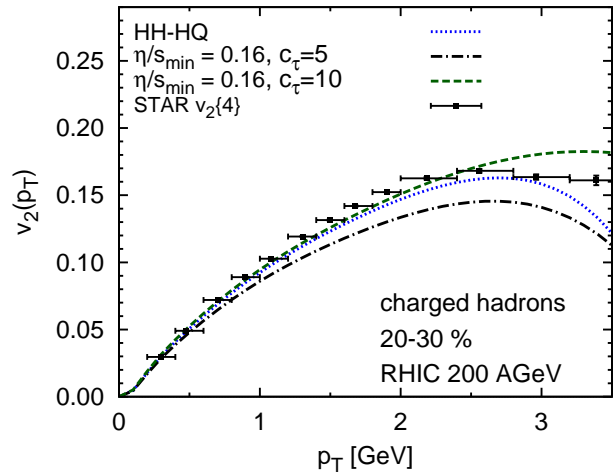} 
% \vspace{-0.3cm} 
\caption{\protect\small (Color online) Charged hadron $v_2(p_T)$ at RHIC, 
with \emph{BCfit} initialization and for different minima of $\eta/s$ and relaxation times.}
\vspace{-0.3cm} 
\label{fig:v2charged_RHIC_NS_2}
\end{figure}

So far we have changed the shear-viscosity parameterization by keeping the 
minimum fixed. In Fig.~\ref{fig:etahighmin} we show the original \emph{HH-HQ} 
parameterization and one where $\eta/s$ around the minimum is twice as large.
Figure~\ref{fig:v2charged_RHIC_NS_2} shows three $v_2(p_T)$ curves for
Au+Au collisions at RHIC: one with the original \emph{HH-HQ} parameterization, 
one with the larger minimum value of $\eta/s$, and the last one
with the same large minimum value of $\eta/s$, but with a larger relaxation
time, i.e., the constant in the relaxation time formula \eqref{eq:relax} is $c_\tau=10$ 
instead of $c_\tau=5$. We note that even a relatively small change in the $\eta/s$
parameterization near the minimum produces quite a visible change in $v_2(p_T)$.
At RHIC, this change can be almost completely compensated 
by adjusting the relaxation time. This shows that in small, rapidly expanding
systems like the one formed in heavy-ion collisions,  
transient effects have considerable influence on the evolution. In other words, the relaxation
time cannot be merely considered as a way to regularize the unstable
Navier-Stokes theory: it has real physical effects that cannot be
completely distinguished from the effects of $\eta/s$. 
In $\sqrt{s_{NN}} = 2.76$ TeV Pb+Pb collisions at LHC,
the effect of changing the minimum or the relaxation 
time is practically the same. 

%%%%%%%%%%%%%%%%%%%%% FIGURE %%%%%%%%%%%%%%%%%%%%%%%%%%%%%%%%
%%%%%%%%%%%%%%%%%%%%% FIGURE %%%%%%%%%%%%%%%%%%%%%%%%%%%%%%%% 

\section{Time evolution of the elliptic flow}

One way to probe the effects of shear viscosity on the elliptic flow
is to calculate the time evolution of the latter. Typically this is done by
calculating the so-called momentum-space anisotropy from the
energy-momentum tensor, 
\begin{equation}
 \varepsilon_p = \frac{\left< T^{xx} - T^{yy}\right>}{\left< T^{xx} + T^{yy}\right>},
\end{equation}
where the $\left< \cdots \right>$ denotes the average over the 
transverse plane. The problem is, however, that one cannot make a direct
connection of $\varepsilon_p$ to the actual value of $v_2$ obtained from
the decoupling procedure. Also, this way of studying the time evolution does not take
into account that, at fixed time, part of the matter is already decoupled,
i.e., the average over the transverse plane includes also matter that
is outside the decoupling surface.
%%%%%%%%%%%%%%%%%%%%% FIGURE %%%%%%%%%%%%%%%%%%%%%%%%%%%%%%%%
\begin{figure}[bht] 
% \vspace{-0.5cm} 
\includegraphics[width=8.0cm]{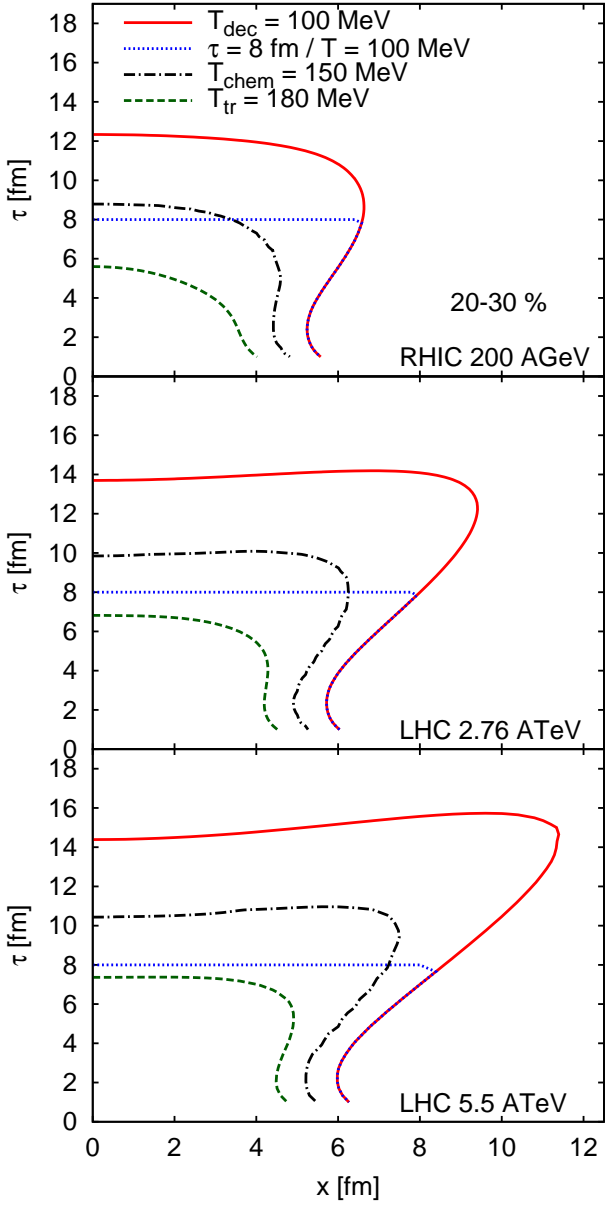} 
% \vspace{-0.3cm} 
\caption{\protect\small (Color online) 
Constant-temperature hypersurfaces at decoupling ($T_{\rm dec} = 100$ MeV), 
chemical freeze-out ($T_{\rm chem} = 150$ MeV), and at the minimum of 
$\eta/s$ ($T_{tr} = 180$ MeV) at different collision energies. 
Also, examples of surfaces that are used in the calculation of 
the time evolution of $v_2$ are shown (dotted lines).}
% \vspace{-0.3cm} 
\label{fig:decs}
\end{figure} 
%%%%%%%%%%%%%%%%%%%%% FIGURE %%%%%%%%%%%%%%%%%%%%%%%%%%%%%%%% 

To overcome these two shortcomings of $\varepsilon_p$, we instead calculate 
the $v_2$ of pions from a constant-time hypersurface that is 
connected smoothly to a constant-temperature hypersurface at the edge 
of the fireball, see Fig.~\ref{fig:decs} for examples of such 
hypersurfaces. Although, the pions do not exist as
real particles before hadronization, the advantage is that the
final $v_2$ we obtain matches the one of thermal pions from the full
decoupling calculation. 

%%%%%%%%%%%%%%%%%%%%% FIGURE %%%%%%%%%%%%%%%%%%%%%%%%%%%%%%%%
\begin{figure}[bht] 
% \vspace{-0.5cm} 
\includegraphics[width=8.0cm]{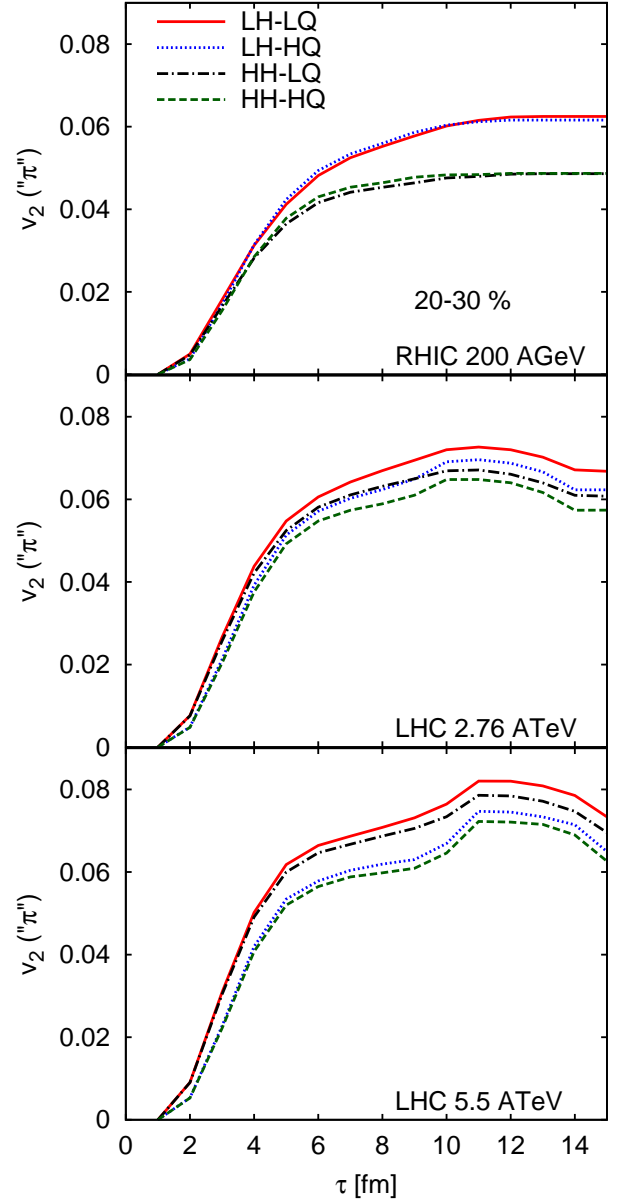} 
% \includegraphics[width=8.0cm]{figs/v2timevo_RHIC.eps} 
% \vspace{-0.3cm} 
\caption{\protect\small (Color online) Time evolution of $v_2$ at different
collision energies.}
% \vspace{-0.3cm} 
\label{fig:v2timevo}
\end{figure} 
%%%%%%%%%%%%%%%%%%%%% FIGURE %%%%%%%%%%%%%%%%%%%%%%%%%%%%%%%% 

Figure~\ref{fig:v2timevo} shows the time evolution of $v_2$ 
in Au+Au collisions at RHIC, in $\sqrt{s_{NN}} = 2.76$ TeV 
Pb+Pb collisions at LHC, and in $\sqrt{s_{NN}} = 5.5$ TeV Pb+Pb 
collisions at LHC. In all cases, the evolution
is calculated in the $20-30$ \% centrality class.
These results confirm our earlier conjecture: 
at RHIC, the different $\eta/s$ parameterizations 
create very little difference in the elliptic flow
in the early stages of the
collision, while at later stages the suppression due to the hadronic
viscosity takes over and groups the $v_2$ curves according to
the hadronic viscosity. At the intermediate LHC energy the 
impact of the QGP viscosity is larger, and the final $v_2$ still has
a memory of this difference. The hadronic viscosity
has a similar impact on $v_2$ as the QGP viscosity. At the highest LHC energy
the hadronic suppression is small and the effect of the QGP viscosity 
clearly dominates the grouping of the $v_2$ curves. Interestingly, both 
LHC evolutions show an increase of $v_2$ around $\tau = 10$ fm/c.
This is when the system is going through the chemical decoupling stage.
In the chemically frozen system $v_2$ tends to increase more
rapidly than in chemical equilibrium~\cite{Huovinen:2007xh, Hiranoetal}. At RHIC,
the chemical decoupling happens earlier, and also the
hadronic suppression is stronger, and the increase in 
$v_2$ is washed out.

\section{Probing the effects of a temperature-dependent \boldmath $\eta/s$ on the $v_n$'s}
\label{probing}

\begin{figure}[bht] 
% \vspace{-0.5cm} 
\includegraphics[width=8.0cm]{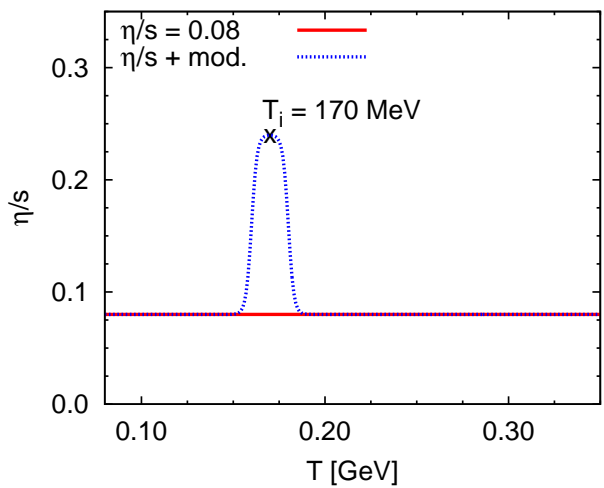} 
% \vspace{-0.3cm} 
\caption{\protect\small (Color online) Shear viscosity with a modified temperature
dependence.}
% \vspace{-0.3cm} 
\label{fig:etapers_shift}
\end{figure}

In this section, we try to probe the effects of a temperature-dependent $\eta/s$
on the azimuthal asymmetries in a more detailed way. To this end, we introduce 
a modified $\eta/s$. Our baseline is a constant $\eta/s|_{c} = 0.08$ that we then 
modify near some temperature $T_{i}$ according to 
\begin{equation}
 \frac{\eta}{s}(T) = \frac{\eta}{s}\Big|_{c}\left[1 + 2 \left(\exp{\left( \frac{|T - T_{i}| - \delta T}{\Delta}\right) } + 1\right)^{-1}  \right],
\label{eq:etapers_mod}
\end{equation}
where the parameters are taken to be $\delta T = 10$ MeV and $\Delta = 1.5$ MeV.
One example of this $\eta/s$ parameterization is shown in Fig.~\ref{fig:etapers_shift}.
We note that, although we use smooth initial conditions from the optical Glauber
model, we still get non-zero $v_n$ for all even $n$. Although these are much
smaller than the ones obtained with the fluctuations included, we can still probe
the effects of viscosity on these coefficients. By changing the temperature $T_{i}$ and 
comparing the simulations with a constant $\eta/s$ we can find the temperature 
regions where $v_2$ or $v_4$ are most sensitive to changes of $\eta/s$ at 
different collision energies. 
%%%%%%%%%%%%%%%%%%%%% FIGURE %%%%%%%%%%%%%%%%%%%%%%%%%%%%%%%%
%%%%%%%%%%%%%%%%%%%%% FIGURE %%%%%%%%%%%%%%%%%%%%%%%%%%%%%%%% 

\begin{figure}[bht] 
% \vspace{-0.5cm} 
\includegraphics[width=8.0cm]{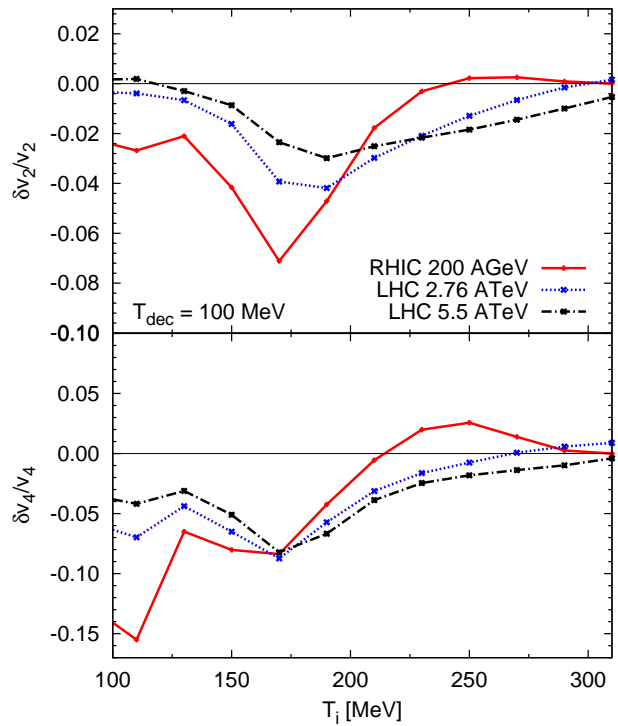} 
% \vspace{-0.3cm} 
\caption{\protect\small (Color online) Effects of modified $\eta/s$ on $v_2$ and $v_4$.}
% \vspace{-0.3cm} 
\label{fig:vnT}
\end{figure}

\begin{figure}[bht] 
% \vspace{-0.5cm} 
\includegraphics[width=8.0cm]{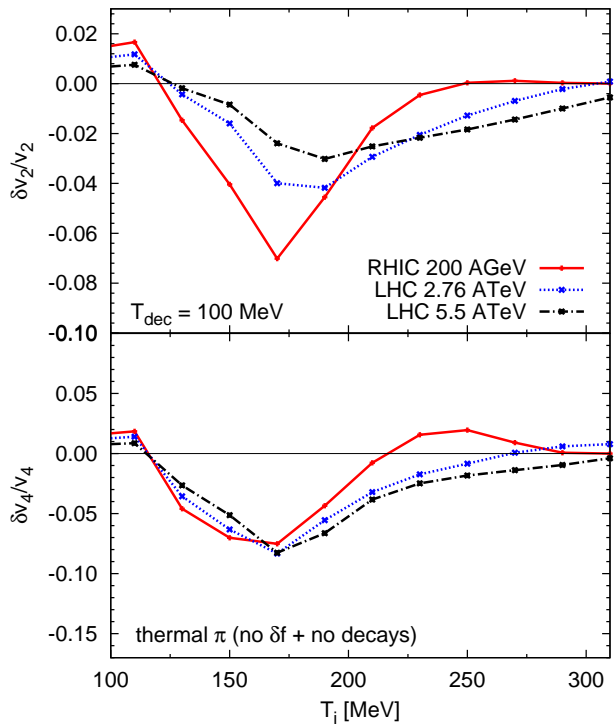} 
% \vspace{-0.3cm} 
\caption{\protect\small (Color online) Same as Fig.~\ref{fig:vnT}, but without the $\delta f$
contribution.}
% \vspace{-0.3cm} 
\label{fig:vnT_thermal}
\end{figure}

Figure \ref{fig:vnT} shows the results for $v_2$ and $v_4$ in the $20-30$~\% centrality class
for RHIC and for both LHC energies considered earlier. We plot the relative difference
$\delta v_n /v_n$, where $\delta v_n = v_n(\eta/s(T)) - v_n(\eta/s|_c)$. 
Each point in the figure corresponds to a
different calculation, with a different value of $T_{i}$ in Eq.~\eqref{eq:etapers_mod}.
Similarly, Fig.~\ref{fig:vnT_thermal} shows the same result, but without the $\delta f$ contribution
to the freeze-out.
%%%%%%%%%%%%%%%%%%%%% FIGURE %%%%%%%%%%%%%%%%%%%%%%%%%%%%%%%%
%%%%%%%%%%%%%%%%%%%%% FIGURE %%%%%%%%%%%%%%%%%%%%%%%%%%%%%%%% 
%%%%%%%%%%%%%%%%%%%%% FIGURE %%%%%%%%%%%%%%%%%%%%%%%%%%%%%%%%
%%%%%%%%%%%%%%%%%%%%% FIGURE %%%%%%%%%%%%%%%%%%%%%%%%%%%%%%%% 

The viscosity can affect $v_n$ in two ways: by changing the space-time evolution
of the integrated quantities like the energy density, or by changing the local 
particle-distribution function at freeze-out. With our small baseline viscosity the effect
on the local distribution function is quickly washed out during the evolution
below the temperature $T_{i}$. Therefore, in these simulations, in most of the temperature
points, the change in $\eta/s$ affects $v_n$ through the space-time evolution, except
at the lowest-temperature point $T_{i} = 110$ MeV, where the peak in $\eta/s$ is close
to the freeze-out temperature $T_{\rm dec} = 100$ MeV.
If we exclude the lowest temperature point in $v_4$ at RHIC, 
we can read off from the figures that the temperature region where viscosity affects 
both $v_2$ and $v_4$ most is around the transition region $T \sim 150 \ldots 200$ MeV.
For $v_2$ this temperature region shifts slightly towards higher temperatures
with increasing collision energy, while for $v_4$ the temperature where the effect
is maximal is practically unchanged. Other than this, the overall 
behavior of $v_2$ and $v_4$ is quite similar. At high temperatures, the
effect of $\eta/s$ increases with increasing collision energy, while at
low temperatures the viscous suppression decreases with increasing collision
energy, which is most notable for the $T_{i} = 110$ MeV point where the viscosity
effects on the freeze-out distribution are strongest. 

For $v_2$ we observed earlier that the suppression due to the hadronic viscosity 
practically vanishes at the highest-energy LHC collisions. This is again confirmed
in Fig.~\ref{fig:vnT}. This is, however, not true for higher harmonics. For $v_4$ there
is still a significant contribution from hadronic viscosity at the full LHC energy.
In this sense, higher harmonics do not give direct access to the 
high-temperature viscosity, but can rather help in constraining the hadronic dynamics
and viscosity as well as the correct form of $\delta f$. This is also important since
the hadronic evolution always tends to shadow the effects of the properties 
of the high-temperature matter.

\section{Conclusions}

We have studied the effects of a temperature-dependent $\eta/s$ on
the azimuthal asymmetries of hadron transverse momentum spectra.
We found earlier \cite{Niemi:2011ix} that the viscous suppression of the elliptic flow 
is dominated by the hadronic viscosity in $\sqrt{s_{NN}} = 200$ GeV Au+Au collisions 
at RHIC, while in Pb+Pb collisions at the full LHC energy $\sqrt{s_{NN}} = 5.5$ TeV
the suppression is mostly due to the high-temperature shear viscosity.
In this work we have supplemented these earlier studies with more details.

First, we found that the suppression of the elliptic flow due
to the shear viscosity becomes more important in more peripheral 
collisions. At least in our set-up, for RHIC energies a temperature-dependent shear 
viscosity improves the centrality dependence of the elliptic flow 
compared to the data, similarly to what was found in the hybrid 
approach of Ref.~\cite{SongBassetal}.
With a constant $\eta/s = 0.08$ and with the \emph{GLmix} 
initialization, the measured $v_2(p_T)$ is reproduced in the most central 
collisions, but the calculations give a too large elliptic flow for
peripheral collisions. However, with the \emph{BCfit} initialization
the elliptic flow in the most central collisions is reproduced with
a temperature-dependent viscosity, and also the centrality
dependence is reproduced down to the $30-40$ \% centrality class.
Similarly, in Pb+Pb collisions at LHC both a temperature-dependent 
hadronic $\eta/s$ as well as an increasing $\eta/s$
in the high-temperature phase help in reproducing the centrality
dependence. Although there are lots of uncertainties associated with the decoupling
and the initial state, at RHIC the centrality dependence of $v_2(p_T)$ may
give access to the temperature dependence of $\eta/s$ in hadronic matter. 

Furthermore, we have studied the effects of a
temperature-dependent $\eta/s$ in a more detailed way.
We found that for a given collision energy both
$v_2$ and $v_4$ are most sensitive to the shear viscosity
near the transition temperature, i.e., $T \sim 150-200$ MeV.
For $v_2$, this region moves slightly to higher temperature and widens with increasing 
collision energy, while for $v_4$ it remains practically unchanged. 
Other than that, the dependence of $v_2$ and $v_4$ on 
$\eta/s$ is similar with increasing collision energy: the
effect of the hadronic viscosity decreases and the effect of the high-temperature 
viscosity increases.

For $v_2$ the effect of $\delta f$ almost vanishes at the highest collision
energies, but for $v_4$ it always remains significant. At RHIC the $\delta f$
corrections clearly dominate $v_4$, and even at the highest collision
energies
this effect is comparable to the effects due to the modified space-time evolution.
In this sense, higher harmonics give access to the $\delta f$ 
corrections and the hadronic viscosity rather than the high-temperature viscosity.

\section*{Acknowledgement}

This work was supported by the Helmholtz International Center for FAIR within the
framework of the LOEWE program launched by the State of Hesse. 
G.S.D., P.H., E.M., and D.H.R.\ acknowledge the hospitality of 
the Department of Physics of Jyv\"askyl\"a University where part of this
work was done. The work of H.N.\ was supported by the Extreme Matter Institute (EMMI), that of
P.H.\ by BMBF under contract no.\ 06FY9092, and that of
E.M.\ by the Hungarian National Development Agency OTKA/NF\"{U} 81655.

\end{document}